\documentclass{ws-mpla}

\begin{document}
\def\beq{\begin{equation}}
\def\enq{\end{equation}}
\def\beqa{\begin{eqnarray}}
\def\enqa{\end{eqnarray}}
\def\nnb{\nonumber}
\newcommand{\rag}{\rangle}
\newcommand{\lag}{\langle}
\def\MeV{\nobreak\,\mbox{MeV}}
\def\GeV{\nobreak\,\mbox{GeV}}
\def\pli{p^\prime}
\def\ql{{p^\prime}^2}
\def\mli{{M^\prime}^2}
\def\muli{\mu^\prime}
\def\nuli{\nu^\prime}
\def\ali{\alpha^\prime}
\def\nn{\nonumber}
\def\lb{\label}
\def\qq{\lag\bar{q}q\rag}

\markboth{M. Nielsen}
{Charged Exotic Charmonium States} 

\catchline{}{}{}{}{}

\title{CHARGED EXOTIC CHARMONIUM STATES}

\author{\footnotesize MARINA NIELSEN}

\address{Instituto de F\'{\i}sica, Universidade de S\~{a}o Paulo\\
C.P. 66318, 05389-970 S\~{a}o Paulo, S\~ao Paulo, Brasil\\
mnielsen@if.usp.br}

\author{FERNANDO S. NAVARRA}

\address{Instituto de F\'{\i}sica, Universidade de S\~{a}o Paulo\\
C.P. 66318, 05389-970 S\~{a}o Paulo, S\~ao Paulo, Brasil\\
navarra@if.usp.br}

\maketitle

\pub{Received (Day Month Year)}{Revised (Day Month Year)}

\begin{abstract}

In this short review we present and discuss all the experimental information 
about the charged exotic charmonium states, which have been observed over the 
last five years. We try to understand their properties such as masses and decay 
widths with QCD sum rules. We describe this method, show the results and compare 
them with the experimental data and with other theoretical approaches.

\end{abstract}

\ccode{PACS Nos.: 11.55.Hx, 12.38.Lg , 12.39.-Mk}

\section{Introduction}	

Since its first observation in 2003 by the  Belle Collaboration
\cite{bellex}, the $X(3872)$ has attracted the interest of all the hadronic 
community. It is the most well studied state among the new charmonium states
and has been confirmed by  five collaborations: CDF \cite{cdfx},
D0 \cite{d0x}, BaBar \cite{babarx}, LHCb \cite{lhcbx} and CMS \cite{cmsx}.
There is little doubt in the
community that the $X(3872)$ structure is more complex than  just a  
$c\bar{c}$ state. Besides the $X(3872)$ the other recently observed
charmonium states that clearly have a more complex structure than
 $c\bar{c}$ are the charged states. Up to now there are some experimental
evidences for seven charged states, which are shown in Table 1.

\begin{table}[h]
\tbl{Charged exotic charmonium states}
{\begin{tabular}{@{}ccccc@{}} \toprule
State (Mass) & Experiment (Year) & $J^{P}$ & Decay Mode & Ref.\\ 
\colrule
$Z^+(4430)$    & BELLE (2008)  &     $1^{+}$ &   $B^+\to K\psi'\pi^+$ & 
\refcite{bellez}\\
$Z^+_1 (4050)$ & BELLE (2008)  &           ? &  $\bar{B}^0\to K^-\pi^+\chi_{c1}$
& \refcite{belle3}\\
$Z^+_2 (4250)$ & BELLE (2008)  &           ? & $\bar{B}^0\to K^-\pi^+\chi_{c1}$  
& \refcite{belle3} \\
$Z^+_c (3900)$ & BESIII (2013) &  $1^+$  &  $Y(4260)\to (J/\psi \pi^+) \pi^-$  &
\refcite{Ablikim:2013mio}\\
$Z^+_c (4025)$ & BESIII (2013) &  ?  &  $e^+e^-\to (D^*\bar{D}^*)^\pm \pi^\mp$  
& \refcite{Ablikim:2013emm}\\
$Z^+_c (4020)$ & BESIII (2013) &  ?  &  $e^+e^-\to (\pi^+h_c) \pi^- $  
& \refcite{Ablikim:2013wzq}\\
$Z^+_c (3885)$ & BESIII (2013) &  ?      &  $e^+e^-\to (D\bar{D}^*)^\pm 
\pi^\mp$  & \refcite{Ablikim:2013xfr}\\ 
\botrule
\end{tabular}\label{ta1} }
\end{table}

The first charged charmonium state, the $Z^+(4430)$, was observed by the Belle 
Collaboration in 2008, produced in $B^+\to K (\psi'\pi^+)$ \cite{bellez}. However 
the Babar Collaboration  \cite{babarz}  searched for the $Z^-(4430)$ signature in
four decay modes and concluded that there is no significant evidence for a 
signal peak  in any of these processes. Very recently the
Belle Collaboration has confirmed the $Z^+(4430)$ observation and has 
determined the  preferred assignment of the quantum numbers to be 
$J^{P} = 1^{+1}$ \cite{bellez2}. Curiously, there are no reports of a $Z^+$ 
signal in the $J/\psi\pi^+$ decay channel.

The $Z^+(4430)$ observation motivated further studies of other $\bar{B}^0 $ decays.
The Belle Collaboration has  reported the observation of two 
resonance-like structures, called  $Z_1^+(4050)$ and $Z_2^+(4250)$, 
in the exclusive process  $\bar{B}^0\to K^-\pi^+\chi_{c1}$, in the $\pi^+
\chi_{c1}$ mass distribution \cite{belle3}. 
Once again the BaBar colaboration did not confirm these observations 
\cite{Lees:2011ik}.

After these non confirmations, it was with great excitement
that the hadron community heard about the observation of the $Z_c^+(3900)$. 
The $Z_c^+(3900)$ was first observed by the BESIII 
collaboration in the $(\pi^\pm J/\psi)$ mass spectrum 
of the $Y(4260)\to J/\psi\pi^+\pi^-$ decay channel \cite{Ablikim:2013mio}.
This structure, was also observed at the same time by the  
Belle collaboration \cite{Liu:2013dau} and was confirmed by the  authors of 
Ref. ~[\refcite{Xiao:2013iha}] using  CLEO-c data.

Soon after the $Z_c^+(3900)$ observation, the BESIII related the observation 
of other three charges states: $Z_c^+(4025)$ \cite{Ablikim:2013emm}, 
$Z_c^+(4020)$ \cite{Ablikim:2013wzq}    and $Z_c^+(3885)$ \cite{Ablikim:2013xfr}.
Up to now it is not clear if the states $Z_c^+(3900)$-$Z_c^+(3885)$
and the states $Z_c^+(4025)$-$Z_c^+(4020)$ are the same states seen in different
decay channels, or if they are independent states. 

All  these charged states can not be  $c\bar{c}$ states and they are
natural candidates for molecular or tetraquark states. 
These exotic states are allowed by the strong interactions, both at the 
fundamental level and at  the effective level, and their absence in the 
experimentally measured spectrum has  always been  a mystery. The theoretical 
tools to address these questions are lattice QCD, chiral perturbation theory, 
QCD sum rules (QCDSR), effective lagrangian approaches and quark models. 
For more details we refer the reader to the more comprehensive  
Ref.~[\refcite{nora}] and to the more recent and  also more specific 
Ref.~[\refcite{reliu}] review articles.

In this rapidly evolving field, periodic accounts of the status of theory and 
experiment are needed. There are already several reviews of the recent charmonium 
spectroscopy. The  
present one is focused on the charged states and on the QCDSR approach to them. 
In the next sections we discuss some of these new charmonium states
using the QCDSR approach.

\section{QCD Sum Rules}

The method of the QCDSR is a powerful tool to evaluate the masses and decay 
widths of  hadrons based on first principles. It was first introduced by 
Shifman, Vainshtein and Zakharov \cite{svz} to the study of  mesons, and 
was latter extended to baryons by Ioffe \cite{io1} and Chung {\it et al.} 
\cite{dosch,dosch2}. Since then the QCDSR technique has been applied to study 
numerous hadronic properties with various flavor content and has been discussed 
in many reviews \cite{rry,SNB,SNB2,col,review}
emphasizing different aspects of the method.
The method is based on identities between two- or three-point 
correlation functions, which connect hadronic observables with QCD fundamental
parameters, such as quark masses, the strong coupling constant, and 
quantities which characterize the QCD vacuum, i.e., the condensates. 
The correlation function is of a dual nature: it represents 
a  quark-antiquark fluctuation 
for short distances (or large momentum) and can be treated in perturbative
 QCD, while at large distances (or small momentum) it can be related to 
hadronic observables.
The sum rule calculations are based on the assumption that in some range of 
momentum both descriptions are equivalent. One, thus, proceeds by calculating 
the correlation function
for both cases and by eventually equating them to obtain  information on the
 properties of the hadrons.

In principle, QCDSR allow first-principle calculations.
In practice, however, in order to extract  results, it is
necessary to make expansions, truncations, and other approximations
that may reduce the power of the formalism and introduce large errors. 
However, if one can find ways to control these errors, the method can provide
important informations about the structure of the hadrons.

\subsection{Hadron masses}

The QCD sum rule calculations of the mass of a hadronic state  are based on
the correlator of two hadronic currents. A generic two-point
correlation function is given by
\beq
\Pi(q)\equiv i\int d^4 x\, e^{iq\cdot x}
\lag{0}| T [j(x)j^\dagger(0)]|0\rag\ ,
\label{cor}
\enq
where $j(x)$ is a current with the quantum numbers of the hadron we want to
study. In the QCDSR approach the correlation function
is evaluated in two different ways: at the quark level in terms of quark 
and gluon fields and at the hadronic level
introducing hadron characteristics such as the mass and the coupling
of the hadronic state to the current $j(x)$. 

The  hadronic side, or phenomenological side of the sum rule is evaluated  by 
writing a dispersion relation to the correlator in Eq.~(\ref{cor}):
\beq
\Pi^{phen}(q^2)=-\int ds\, {\rho(s)\over q^2-s+i\epsilon}\,+\,\cdots\,,
\label{disp}
\enq
where $\rho$ is the spectral density given by the absorptive part of 
the correlator and the dots represent subtraction terms. 

Since
the current $j~(j^\dagger)$ is an operator that
annihilates (creates) all hadronic states that have the same quantum
numbers as $j$, $\Pi(q)$ contains information about
all these hadronic states, including the low mass hadron of interest.
In order  the QCDSR technique to be useful, one must parameterize
$\rho(s)$ with a small number of parameters.  In general one parameterizes 
the spectral density as a single sharp
pole representing the lowest resonance of mass $m$, plus a smooth continuum 
representing higher mass states:
\beq
\rho(s)=\lambda^2\delta(s-m^2) +\rho_{cont}(s)\,,
\label{den}
\enq
where $\lambda$ gives the coupling of the current with
the low mass hadron, $H$: $\lag 0 |j|H\rag =\lambda$. With this ansatz
the phenomenological side of the sum rule becomes:
\beq
\Pi^{phen}(q^2)=-{\lambda^2\over q^2-m^2}\,-\,
\int_{s_{min}}^{\infty} ds\, {\rho_{cont}(s)\over q^2-s+i\epsilon}\,+\,\cdots\,,
\label{phen}
\enq

In the QCD side, or OPE side the correlation function is evaluated by using
the Wilson's operator product expansion (OPE) \cite{ope}:
\beq
\Pi^{OPE}(q)=\sum_nC_n(Q^2)\hat{O}_n\;,
\label{cope}
\enq
where the set $\{\hat{O}_n\}$ includes all local gauge invariant
operators expressible in terms of the gluon fields and the fields of light
quarks, which are represented  in the form of vacuum condensates. The 
lowest dimension condensates are the
quark condensate of dimension three: $\hat{O}_3=\langle\bar{q}q\rangle$,
and the gluon condensate of dimension four: $\hat{O}_4=
\langle g^2 G^2\rangle$. The lowest-dimension
operator with $n=0$ is the unit operator associated with the perturbative
contribution.

For non exotic mesons, {\it i.e.} normal 
quark-antiquark states, such as   $\rho$ and $J/\psi$, the contributions of 
condensates with dimension higher than four are suppressed by large powers of 
$1/Q^2$. Therefore, the expansion in Eq.~(\ref{cope}) can be safely truncated 
after dimension four condensates, even at intermediate values of 
$Q^2~(\sim1\GeV^2)$. However, for molecular or tetraquark states, higher 
dimension condensates like the dimension 
five mixed-condensate: $\hat{O}_5=\lag\bar{q}g\sigma.Gq\rag$, the dimension 
six four-quark condensate:  $\hat{O}_6=\lag\bar{q}q\bar{q}q\rag$ and even the
dimension eight quark condensate times the mixed-condensate:
$\hat{O}_8=\lag\bar{q}q\bar{q}g\sigma.Gq\rag$, can play an important role.
The three-gluon condensate of dimension-six: $\hat{O}_6=\lag g^3 G^3\rag$ 
can be safely neglected, since it is suppressed by the loop factor 
$1/16\pi^2$. 

The precise evaluation of the  $D=6$, $\hat{O}_6$, and $D=8$, $\hat{O}_8$,
condensates require a involved analysis including a non-trivial choice of 
factorization scheme \cite{BAGAN}. Therefore, in our calculations we 
assume that  their vacuum saturation values are given by:
\beq
\lag\bar{q}q\bar{q}q\rag=\lag\bar{q}q\rag^2,\;\;\;\lag\bar{q}q\bar{q}
g\sigma.Gq\rag=\lag\bar{q}q\rag\lag\bar{q}g\sigma.Gq\rag.
\enq

The OPE side can also be written in terms of a dispersion relation as:
\beq
\Pi^{OPE}(q^2)=-\int_{s_{min}}^{\infty} ds\, {\rho^{OPE}(s)\over q^2-s+i\epsilon}
\,+\,\cdots\,,
\label{ope}
\enq
where 
\beq
\rho^{OPE}(s)={1\over\pi}Im[\Pi^{OPE}(s)]\;.
\enq

To keep the number of parameters as small as possible, in general
in the QCDSR approach one assumes that the continuum contribution to the
spectral density, $\rho_{cont}(s)$ in Eq.~(\ref{phen}), vanishes bellow a
certain continuum threshold $s_0$. Above this threshold one uses the ansatz
\beq
\rho_{cont}(s)=\rho^{OPE}(s)\Theta(s-s_0)\;.
\label{cont}
\enq
Using Eq.~(\ref{cont}) in Eq.~(\ref{phen}) we get
\beq
\Pi^{phen}(q^2)=-{\lambda^2\over q^2-m^2}\,-\,
\int_{s_{0}}^{\infty} ds\, {\rho^{OPE}(s)\over q^2-s+i\epsilon}\,+\,\cdots\,,
\label{phen2}
\enq

To improve the matching of the two descriptions of the correlator
one applies the Borel transformation. The Borel transformation removes the 
subtraction terms in the dispersion relation, and
exponentially suppresses the contribution from excited resonances and
continuum states in the phenomenological side. In the OPE side the Borel
transformation suppresses the contribution from higher dimension condensates
by a factorial term.

After performing  a Borel transform on both sides of the sum rule, and
transferring the continuum contribution to the OPE side, the sum rule
can be written as
\beq \lambda^2e^{-m^2/M^2}=\int_{s_{min}}^{s_0}ds~
e^{-s/M^2}~\rho^{OPE}(s)\;. \label{sr} 
\enq

A good sum rule is obtained in the case that one can find a range of $M^2$, 
called Borel window, in which the two sides have a good overlap and information 
on the lowest resonance can be extracted. To determine the allowed Borel window, 
one analyses the OPE convergence and the pole contribution: the minimum value of 
the Borel mass is fixed by considering the convergence of the OPE, and the maximum 
value of the Borel mass is determined by imposing the condition that the pole 
contribution must be bigger than the continuum contribution.

The mass of the hadronic state, $m$, can be obtained by taking the derivative of 
Eq.~(\ref{sr}) with respect to $1/M^2$, and dividing the result by 
Eq.~(\ref{sr}):
\beq
m^2={\int_{s_{min}}^{s_0}ds ~e^{-s/M^2}~s~\rho^{OPE}(s)\over\int_{s_{min}}^{
s_0}ds ~e^{-s/M^2}~\rho^{OPE}(s)}\;.
\label{m2}
\enq

Using the formalism described above we can compute the masses of the new states. 
A compilation of results of the states discussed here is shown in Table 2. These 
numbers will be discussed in detail in the next sections.
\begin{table}[h]
\tbl{Masses obtained with QCDSR}
{\begin{tabular}{@{}ccccc@{}} \toprule
State  & $J^{PC}$  & Current & Mass & Ref.\\
\colrule
$X(3872)$     & $1^{++}$  & Tetraquark  &   $(3.92 \pm 0.13)$ GeV & 
\refcite{x3872}\\
$X(3872)$     & $1^{++}$  & $D\bar{D^*}$ Molecule    &   $(3.87 \pm 0.07)$ GeV & 
\refcite{korea}\\
$Z^+(4430)  $ & $0^-$     & $D^* \bar{D_1}$ Molecule  &   $(4.40 \pm 0.10)$ GeV & 
\refcite{Lee:2007gs}\\
$Z^+(4430)  $ & $0^-$     & Tetraquark  &   $(4.52 \pm 0.09)$ GeV & 
\refcite{Bracco:2008jj}\\
$Z^+(4430)  $ & $1^-$     & Tetraquark  &   $(4.84 \pm 0.14)$ GeV & 
\refcite{Bracco:2008jj}\\
$Z^+_1 (4020)$ & $0^+ $   & $D^* \bar{D^*}$ Molecule & $(4.15 \pm 0.12) $ GeV  & 
\refcite{morita}\\
$Z^+_2 (4250)$ & $1^- $   & $D_1 \bar{D}$   Molecule & $(4.19 \pm 0.22) $ GeV  & 
\refcite{morita}\\
$Z_c^+(3930) $ & $1^+$    & Tetraquark  &   $(3.92 \pm 0.13)$ GeV & 
\refcite{Dias:2013xfa}\\
$Z_c^+(4025) $ & $1^+$    & $D^* \bar{D^*}$ Molecule &   $(3.950 \pm 0.105)$ GeV & 
\refcite{khetonn}\\
$Z_c^+(4025) $ & $2^+$    & $D^* \bar{D^*}$ Molecule &   $(3.946 \pm 0.104)$ GeV & 
\refcite{khetonn}\\
 \botrule
\end{tabular}\label{ta2} }
\end{table}

\subsection{Hadron decay widths}

The QCD sum rule calculations for the coupling constant in a hadronic vertex
are based on the correlator of three hadronic currents. 
A generic three-point correlation function  associated with a  vertex of 
three mesons $M_1$, $M_2$ and $M_3$ is given by
\beq
\Gamma (p,\pli,q) = \int d^4x \, d^4y \;\; e^{i\pli\cdot x} 
\, e^{-iq\cdot y}  \langle 0|T\{j_{3}(x) j_{2}^{\dagger}(y) 
 j_{1}^{\dagger}(0)\}|0\rangle\, 
\label{corr} 
\enq
where $q=\pli-p$ and the current $j_i$ represents states with the quantum numbers 
of the meson $i$. As in the case of the two-point correlation function, the 
function 
in Eq.~(\ref{corr}) is evaluated in two ways. In the OPE side we consider that 
the currents are composed by quarks and we use the Wilson's OPE to 
evaluate the correlation function. In the phenomenological side, 
we  insert, in Eq.(\ref{corr}), intermediate states for the mesons $M_1$, $M_2$ 
and $M_3$. We then write the correlation function in terms of the coupling
of these mesons with the corresponding currents, and in terms of the form factor, 
$g_{M_1M_2M_3}(q^2)$, in the hadronic vertex, which is defined by the 
generalization of the on-mass-shell matrix element, $\lag M_3M_2|M_1\rag$,
for an off-shell $M_2$ meson:
\beq
\lag M_3(\pli) M_2(q)|M_1(p)\rag=g_{M_1M_2M_3}(q^2)f_{M_{1},p} f_{M_{2},\pli} 
f_{M_{3},q},
\label{coup}
\enq
which can be extracted from the effective Lagrangian that describes the
coupling between these three mesons. In Eq.~(\ref{coup}) the functions
$f_{M_{i},k}$ are obained from the Lagrangian and are related with the
quantum numbers of the meson $M_i$. After evaluating both sides  separately, 
we equate one description with the other and we can extract the
form factor from  the sum rule. 

The coupling constant is defined as the value of the form factor at the meson 
pole: $Q^2=-m_2^2$, where $m_2$ is the mass of the meson $M_2$ that was off-shell. 
Very often, in order to determine the coupling constant we have to extrapolate 
the QCDSR results to a $Q^2$ region where the sum rules are no longer valid 
(since the QCDSR results are valid in the deep Euclidian region). To do that, 
in general, we parametrize the QCDSR results through a analytical form, like
a monopole or an exponential function. For more details we refer the reader 
to Ref.~[\refcite{Bracco:2011pg}].

\section{ $X(3872)$}

The $X(3872)$ was first observed by Belle  
collaboration in 2003 in the decay $B^+\!\rightarrow\!X(3872)K^+\rightarrow\!J/
\psi\pi^+\pi^- K^+$ \cite{bellex}, and has been confirmed by other five 
collaborations \cite{cdfx,d0x,babarx,lhcbx,cmsx}.
The current world average mass is $m_X=(3871.68 \pm0.17)$ MeV
and its total width is less than $1.2$ MeV \cite{pdg}. The LHCb collaboration 
determined $J^{PC} = 1^{++}$ quantum numbers with more than 8$\sigma$ 
significance \cite{lhcbx2}.

\subsection{Mass}

Calculations using constituent quark models 
give masses for  possible charmonium states, with $J^{PC}=1^{++}$ quantum 
numbers,  which are much bigger than the observed $X(3872)$ mass: 
$2~^3P_1(3990)$ and $3~^3P_1(4290)$ \cite{bg}. These results, together with the 
coincidence between the $X$ mass and the $D^{*0}D^0$ threshold: 
$M(D^{*0}D^0)=(3871.81\pm0.36)\MeV$ \cite{cleo}, inspired the proposal that 
the $X(3872)$ could be a molecular $(D^{*0}\bar{D}^0+\bar{D}^{*0}D^0)$ bound 
state with small binding energy \cite{swanson,swanson2,swanson3,swanson4,swanson5}.

Other interesting possible interpretation of the $X(3872)$, first proposed by
Maiani {\it et al.} \cite{maiani}, is that it could be a tetraquark state
resulting from the binding of a diquark and an antidiquark. 

The fisrt QCDSR calculation of the mass of  the $X(3872)$ considered as a 
$J^{PC}=1^{++}$ tetraquark state was done in Ref.~[\refcite{x3872}]. 
Following this calculation,  a $J^{PC}=1^{++}$, $D^{*}
\bar{D}$ molecular current was considered in Ref.~[\refcite{korea}]. 
The corresponding interpolating currents used in these calculations are: 
\beqa
j^{di}_\mu&=&{i\epsilon_{abc}\epsilon_{dec}\over\sqrt{2}}[(q_a^TC
\gamma_5c_b)(\bar{q}_d\gamma_\mu C\bar{c}_e^T)
+(q_a^TC\gamma_\mu c_b)
(\bar{q}_d\gamma_5C\bar{c}_e^T)]\;,
\label{cur-di}
\enqa
for a tetraquark current, and
\beq
j^{mol}_{\mu} =  {1 \over \sqrt{2}}
\bigg[(\bar{q}_a \gamma_{5} c_a)(\bar{c}_b \gamma_{\mu}  q_b)
-(\bar{q}_a \gamma_{\mu} c_a)(\bar{c}_b \gamma_{5}  q_b)
\bigg],
\lb{cur-mol}
\enq
for a molecular $D\bar{D}^*$ current. In Eqs.~(\ref{cur-di}) and 
(\ref{cur-mol}), $q$ denotes a $u$ or $d$ quark.

In the OPE side, the calculations were done at leading order
in $\alpha_s$ and contributions of condensates up to dimension eight were
included. In both cases it was possible to find a Borel window where the pole
contribution is bigger than the continuum contribution and with a reasonable
OPE convergence. 

The mass obtained in Ref.~[\refcite{x3872}], considering the allowed Borel window
and the uncertaities in the parameters, was $m_X=(3.92\pm0.13)$ GeV
whereas the result for the mass obtained in Ref.~[\refcite{korea}] was 
$m_X=(3.87\pm0.07)$ GeV, as shown in Table 2. 

We see that, in both cases, a good agreement with the experimental mass was 
obtained. Up to now there are  many QCDSR calculations\cite{review} of the  
the mass of  the $X(3872)$ considering different currents and in all cases good 
agreement with the experimental mass is found. Even with a mixed 
charmonium-molecular current the value obtained for the mass does not change 
significantly \cite{x24}.
These calculations only confirm the result presented in 
Ref.~[\refcite{Narison:2010pd}] that shows that the calculation of the mass of a 
given state, in the QCDSR approach, is very insensitive to the choice of the 
current. However, this may not be the case for the decay 
width\cite{Narison:2010pd}.

\subsection{Decay width}

The first  QCDSR calculation of the width of  the $X(3872)$  was done in 
Ref.~[\refcite{width}]. In particular, in Ref.~[\refcite{width}] the $X(3872)$ was
considered as a  tetraquark state described by the current in Eq.~(\ref{cur-di}) 
and a very large decay width was obtained:
$\Gamma(X\to J/\psi\rho\to J/\psi\pi^+\pi^-)=(50\pm15)~\MeV$.
A similar width was obtained in Ref.~[\refcite{x24}] with a molecular current
such as the one in Eq.~(\ref{cur-mol}). Indeed, large partial decay widths are 
expected when the coupling constant is obtained from QCDSR, in the case of 
multiquark states, when the 
initial state contains the same number of valence quarks as the number of 
valence quarks in the final state. An example is  the case of the light 
scalars $\sigma$ and $\kappa$ studied in Ref.~[\refcite{sca}], which widths are of 
the order of 400 MeV. In the case of the $X\to J/\psi\rho$ decay,  the generic 
decay diagram in terms of quarks has two ``petals'', one associated with the 
$J/\psi$ and the other with the $\rho$.
\begin{figure}[h]
\centerline{\epsfig{figure=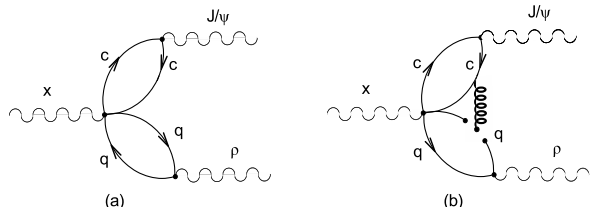,height=30mm}}
\caption{Generic decay diagrams of the $X(3872)\to J/\psi\rho$ decay.}
\label{fig1}
\end{figure} 
 Among the possible diagrams, there are two distinct subsets.  Diagrams with no 
gluon exchange between the petals, as the one shown in Fig.~1(a), and 
therefore, no color exchange between the two final mesons in the decay. If there
is no color exchange, the final state containing two color singlets was
already present in the initial state. In this case the tetraquark had 
a component similar to a $J/\psi-\rho$ molecule. The other subset of diagrams 
is the one where there is a gluon exchange between the 
petals, as the one shown in Fig.~1(b).  
This type of diagram represents the case where the $X$ is a 
genuine four-quark state with a complicated color structure. These diagrams 
are called color-conected (CC). Considering only the CC diagrams in the 
calculation, the decay width obtained in Ref.~[\refcite{width}] was:
\beq
\Gamma_{CC}(X\to J/\psi ~\rho\to J\psi\pi^+\pi^-)=(0.7\pm0.2)~\MeV,
\label{widthcc}
\enq
in a very good agreement with the experimental upper limit.

This procedure may appear somewhat unjustified. However, if the initial state
has a non-trivial color structure only CC diagrams should contribute to the 
calculation. Unfortunately, although the initial tetraquark current  
has a non-trivial color structure, it can be rewritten as a sum 
of molecular type currents with trivial color configuration through a
 Fierz transformation. This is the reason why the diagrams without gluon
exchange between the two ``petals'' survive in the QCDSR calculation.
Therefore, the approach of considering only CC diagrams can be considered as a
form of simulating a real tetraquark state with non-trivial color structure.

Other possible approach to reduce the large widht is to consider the 
$X(3872)$ as a mixture between a $c\bar{c}$ current and a 
molecular current, as done in Ref.~[\refcite{x24}]:
\beq
J_{\mu}(x)= \sin(\alpha) j^{mol}_{\mu}(x) + \cos(\alpha)
j^{2}_{\mu}(x),
\lb{cur24}
\enq
with $j^{mol}_{\mu}(x)$ given in Eq.~(\ref{cur-mol}) and
\beq
j^{2}_{\mu}(x) = {1 \over 6 \sqrt{2}} \qq [\bar{c}_a(x) \gamma_{\mu}
\gamma_5 c_a(x)].
\enq

The necessity of mixing a $c\bar{c}$ component with the $D^0\bar{D}^{*0}$ 
molecule was already pointed out in some works 
\cite{mechao,suzuki,dong,li1}. In particular,
in Ref.~[\refcite{grinstein}], a simulation of the production of a bound 
$D^0\bar{D}^{*0}$ state with binding energy as small as 0.25 MeV, obtained a
cross section of about two orders of magnitude smaller than the prompt
production cross section of the $X(3872)$ observed by the CDF Collaboration.
The authors of Ref.~[\refcite{grinstein}] concluded that $S$-wave resonant 
scattering is unlikely to allow the formation of a loosely bound 
$D^0\bar{D}^{*0}$ molecule in high energy hadron collision. 

As discussed above,
there is no problem in reproducing the experimental mass of the $X(3872)$,
using the current in Eq.~(\ref{cur24}), for a wide range of the mixture 
angle $\alpha$. However, the value of the $XJ/\psi\rho$ coupling constant
and, therefore, the value of the $X\to J/\psi ~(n\pi)$ decay width, is
strongly dependent on this angle. It was shown in Ref.~[\refcite{x24}] that
for a mixing angle $\alpha=9^0\pm4^0$,
it is possible to describe the experimental mass of the $X(3872)$ with
a decay width $\Gamma(X\to J/\psi ~(n\pi))=(9.3\pm6.9)~\MeV$, which is
compatible with the experimental upper limit. Therefore, in a QCDSR
calculation, the $X(3872)$ can be well described basically by a
$c\bar{c}$ current with a small, but fundamental, admixture of molecular
($D\bar{D}^*$) or tetraquark ($[cq][\bar{c}\bar{q}]$) currents.

\section{ $Z^+(4430)$}

This resonance was found by Belle Collaboration in the channel $B^+\to K\psi'\pi^+$
and it was the first charged charmonium  state observed, with mass   
$ M = (4433^{+ 15 + 19}_{-12 -13})$ MeV  and width $\Gamma = (109^{+86 + 74}_{-43 
-56} \pm 18 \pm 30)$ MeV \cite{bellez}.   
Curiously, there is no signal of this resonance in the $J/\psi \pi^+$ channel. 
Since the minimal quark content of this state is $c \bar{c} u \bar{d}$  this  
can only be achieved in a multiquark configuration.  

The Babar Collaboration searched the $Z^- (4430)$ in the four decay modes 
$B^+\to K^0_S \psi'\pi^-$, $B^+\to K^0_S J/\psi \pi^-$, $B^+\to K^+ \psi'\pi^0$ 
and $B^+\to K^+ J/\psi \pi^0$.  No significant evidence of a signal peak was 
found in any of the processes investigated \cite{babarz}. 

Since the  $Z^+(4430)$ mass is close to the $D^* D_1$ threshold, it was suggested that it 
could be an S-wave threshold effect or a $D^* D_1$  molecular state. 
Considering the  $Z^+(4430)$ as a weakly bound $S$-wave $D^*D_1$ molecular 
state, its quantum numbers may be $J^P=0^-,~1^-,~2^-$.  
The $2^-$  assignment is probably suppressed in the $B^+\to Z^+K$ 
decay beacuse of  the small phase space. Other 
possible interpretations are a tetraquark state, a cusp in the $D^* D_1$  
channel, a baryonium state, a radially excited $c \bar{s}$ state and a
hadrocharmonium state \cite{review}. 

There are  QCDSR calculations for the $Z^+(4430)$ assuming that the state could 
have $J^P = 0^-$ or $J^P = 1^-$ quantum numbers \cite{Lee:2007gs,Bracco:2008jj}.  
In the first case the  obtained masses were $m_{mol} (0^-) = (4.40 \pm 0.10)$ GeV 
for a $D^* D_1$ molecular current \cite{Lee:2007gs} and $m_{di} (0^-) = (4.52 \pm 
0.09)$ GeV for a diquark-antidiquark current \cite{Bracco:2008jj}.  
In the second assignment and  for a diquark-antidiquark current the  obtained 
mass was $m_{di} (1^-) = (4.84 \pm 0.14)$ GeV \cite{Bracco:2008jj}.  
These numbers are displayed in Table 2.

From these results the preliminary conclusion, at the time,  was that the 
assignment $J^P = 1^-$ was disfavored and that the configuration $J^P = 0^-$, in 
both molecular and tetraquark states,  would lead to a mass which is in agreement 
with the data. However, a recent reanalysis of the Belle data revealed that 
the favored quantum numbers are $J^P = 1^+$ \cite{bellez2}.  It is important
to mention  that soon after the  $Z^+(4430)$ was first observed, Maiani 
{\it et al.} have suggested that the  $Z^+(4430)$ could be the first radial 
excitation of a charged partner of the $X(3872)$, and therefore, would have 
$J^P = 1^+$ quantum numbers \cite{maiani2}. The existence of a charged partner of 
the $X(3872)$ was first proposed in Ref.~[\refcite{maiani}].

Clearly, in view of the recent experimental reanalysis, if the $Z^+(4430)$ really 
exist, it could be a $\psi' \pi^+$ resonance or a tetraquark excitation, which 
invalidates a QCDSR calculation. 

\section{$Z^+_1 (4050)$ and $Z^+_2 (4250)$} 

After the  observation of the $Z^+(4430)$ other $\bar{B}^0\to K^-\pi^+ (c
\bar{c})$ decays were carefully investigated.  Two resonance-like structures, called $Z_1^+(4050)$ 
and $Z_2^+(4250)$,  were  observed by the Belle Collaboration in the exclusive process  
$\bar{B}^0\to K^-\pi^+\chi_{c1}$, in  the $\pi^+ \chi_{c1}$ mass distribution \cite{belle3}.
The significance of each of the
$\pi^+\chi_{c1}$ structures exceeds 5 $\sigma$ and, since they were observed in
the $\pi^+\chi_{c1}$ channel, they must have the quantum numbers $I^G=1^-$. Also in this case the 
BaBar colaboration did not confirm these observations \cite{Lees:2011ik}.
When fitted with two Breit-Wigner resonance amplitudes, the resonance parameters 
are 
$m_1= ( 4051 \pm 14^{+20}_{-41} )$  MeV, $\Gamma_1 = (82^{+ 21 + 47}_{- 17 - 22})$ 
 MeV, 
$m_2= ( 4248^{+ 44 + 180}_{- 29 - 35} )$ MeV and  $\Gamma_2 = (177^{+ 54 + 316}_{- 
39 - 61})$ MeV. 

Since the masses of the $Z_1^+(4050)$ and $Z_2^+(4250)$ are   close  to the 
$D^* \bar{D^*}(4020)$ and $D_1 \bar{D}(4085)$  thresholds, it is natural to 
interpret these states  as molecular states or threshold effects. However 
calculations using meson exchange models do not agree with each other. In 
Ref.~[\refcite{lium}], a  strong attraction in the  $D^* \bar{D^*}$ with 
$J^P = 0^+$ was found,   while in Ref.~[\refcite{ding}]  the interpretation 
of $Z_1^+(4050)$ as a $D^* \bar{D^*}$ molecule was not favored.  
In any case, it is very difficult to understand a bound molecular
state which mass is above the $D^* \bar{D^*}$ threshold. In Ref.~[\refcite{ding2}]
 the interpretation of $Z_2^+(4250)$ as a $D_1 \bar{D}$ or $D_0 \bar{D^*}$ 
molecule was  disfavored. 

Soon after the observation of these states,  QCDSR were used\cite{morita}  to study 
the $D^{*}\bar{D}^*$ and  $D_1\bar{D}$ molecular states with $I^G J^P=1^-0^+$ 
and $1^-1^-$ respectively. The currents used in both cases were of the type 
of Eq.~(\ref{cur-mol}).   
As shown in Table 2, for the  $D^* \bar{D^*}$ system the  obtained mass was 
$m_{D^*D^*} = (4.15 \pm0.12)$ GeV. Since  the central value of the mass is around 130 MeV 
above the $D^*D^*(4020)$ threshold, we can conclude that there are repulsive interactions 
between the two $D^*$  mesons. Therefore, it is not clear whether  this structure is a 
resonance or not.  For the $D_1 \bar{D}$ system the  obtained mass was $m_{D_1D} = (4.19 \pm0 .22)$ 
GeV. Here, in contrast to the previous case, the central value is around 100 MeV below 
the $D_1 D(4285)$ threshold,  and, considering the errors, consistent with the mass of the $Z_2^+(4250)$  
resonance structure. Therefore, in this case, there seems to be  an atractive interaction 
between the mesons $D_1$ and $D$ and the  molecular interpretation of this state seems more justified.  

QCD sum rules estimate always contain some uncertainties. In the study of the masses of the 
charged $Z$ states, part of the theoretical uncertainty comes from the width of the state. In most cases,
the widht is neglected. In the present case, when the width is included in the  phenomenological side of 
the sum rule,  the mass of the corresponding state increases\cite{leeniel}.  It becomes then  possible 
to obtain a mass  $m_{D_ 1D} = 4.25~\GeV$ with a width  $40\leq\Gamma\leq60~\MeV$. Following the same 
trend, the mass of the $D^*\bar{D}^*$ molecule will  be far  from the $Z_1^+(4050)$ mass. In view of 
these facts, the authors of Ref.~[\refcite{leeniel}] 
concluded that it is possible to describe the $Z_2^+(4250)$  as a 
$D_1\bar{D}$ molecular state with $I^G J^P=1^-1^-$ quantum numbers. They also 
concluded  that the $D^{*}\bar{D}^{*}$ state is probably a virtual state that is not 
related with the $Z_1^+(4050)$ resonance-like structure. Since the $D^*D^*$ threshold $(4020)$ 
is so close to the $Z_1^+(4050)$  mass and  the $\eta^{\prime\prime}_c(3^1S_0)$ mass is 
predicted to be around 4050 MeV, the $Z_1^+(4050)$ is probably only a threshold effect.

\section{$Z^+_c (3900)$}

After the non-confirmed observations of  $Z^+(4430)$,  $Z^+_1 (4050)$ and $Z^+_2 
(4250)$, only seen by Belle,  
the BESIII and Belle collaborations  reported    the observation of a 
charged charmonium-like structure in the $M(\pi^\pm J/\psi)$ mass spectrum of 
the $Y(4260)\to J/\psi\pi^+\pi^-$ decay channel \cite{Ablikim:2013mio,Liu:2013dau}.
The existence of this structure, called $Z_c(3900)$, was promptly confirmed by the 
the  authors of Ref.~[\refcite{Xiao:2013iha}] using  CLEO-c data.   

In most of the theoretical calculations  it is relatively easy to reproduce the 
masses of the states.  In the case of the   $Z_c(3900)$, assuming $SU(2)$  
symmetry, the mass obtained in QCDSR for the $Z_c$  is exactly the same one 
obtained  for the $X(3872)$. As discussed in Sec. {\bf 3.2},
it is, however, much more difficult to reproduce their measured decay
widths. The  $Z_c(3900)$ decay width represents a 
challenge to theorists. While its mass is very close to the $X(3872)$ mass, which may be
considered its isosinglet partner, it has a much larger decay width. Indeed, 
while the $Z_c(3900)$ decay width is in the range $40-50$ MeV, the $X(3872)$ width is  
smaller than  $ 1.2 $ MeV.  

This difference can be attributed to  the fact that the $X(3872)$
may contain a significant $|c \bar{c} \rangle$ component\cite{x24}, which is
absent in the $Z_c(3990)$.  As pointed out in Ref.~[\refcite{zhao}], this would also 
explain why the $Z_c$ has not been observed in $B$ decays. 

According to the experimental observations, the $Z_c(3900)$ decays into $J/\psi \,  \pi^+$
with a relatively large decay width. This is unexpected for a $D^* - \bar{D}$  molecular state, 
in which the distance between the  $D^*$ and the $ \bar{D}$ is large.
This decay must involve the exchange of a charmed meson, which is a short range process and hence 
unlikely to occur in large systems. In Ref.~[\refcite{namit}]  it was shown that, in order to reproduce the
measured width, the effective radius must be $\langle r_{eff} \rangle \simeq 0.4$ fm. This size scale 
is small and pushes the molecular picture to its limit of validity. 
In another work\cite{hammer}, the new state
was treated as a charged $D^* - \bar{D}$ molecule and the authors explored its  
electromagnetic structure, arriving 
at the conclusion that its charge radius is of the order of  $\langle r^2 \rangle \simeq 0.11$ fm$^2$. 
Taking this radius as a measure of the spatial size of the state, 
we conclude that it is more compact than a $J/\psi$, for which
$\langle r^2 \rangle \simeq 0.16$ fm$^2$.  In Ref.~[\refcite{Dias:2013xfa}] the 
combined results of Refs.~[\refcite{namit}] and [\refcite{hammer}] were taken  as 
an indication that the $Z_c$ is a compact object, which may be better understood 
as a quark cluster, such as a tetraquark.  Moreover, the $Z_c(3900)$ was 
interpreted  as the isospin 1 partner of the $X(3278)$, as the charged state 
predicted in Ref.~[\refcite{maiani}]. Therefore, the quantum numbers for the neutral state 
in the isospin multiplet were assumed to be $I^G(J^{PC})=1^+(1^{+-})$.
The interpolating field for $Z_c^+(3900)$ used in Ref.~[\refcite{Dias:2013xfa}] is given
by Eq.~(\ref{cur-di})  with the plus signal changed by a minus 
signal. The three-point QCDSR were used to evaluate 
the coupling constants in the vertices $Z_c^+(3900)J/\psi\pi^+$,  
$Z_c^+(3900)\eta_c\rho^+$,   $Z_c^+(3900) D^+ \bar{D^*}^0 $ 
and   $Z_c^+(3900) \bar{D^0} {D^*}^+ $.  In all cases  only color-connected 
diagrams were considered, since the $Z_c(3900)$ is expected to be a genuine 
tetraquark state with a non-trivial color structure. The obtained couplings, with 
the respective decay widths, are given in Table 3. 
A total width of $\Gamma = (63.0 \pm 18.1)$ MeV was found for the  $Z_c(3900)$,  
in good agreement with the two experimental values:
$\Gamma=(46\pm 22)$ MeV from BESIII \cite{Ablikim:2013mio}, and
$\Gamma=(63\pm35)$ MeV from BELLE \cite{Liu:2013dau}.

\begin{table}[h]
\tbl{ Coupling constants and decay widths in  different channels}
{\begin{tabular} {@{}ccc@{}}   \toprule
Vertex & coupling constant (GeV) & decay width (MeV)\\
\colrule
$Z_c^+(3900)J/\psi\pi^+$ & $3.89\pm0.56$ & $29.1\pm8.2$ \\
 $Z_c^+(3900)\eta_c\rho^+$ & $4.85 \pm 0.81$ & $27.5\pm8.5$ \\
 $Z_c^+(3900) D^+ \bar{D^*}^0 $ & $2.5 \pm 0.3$ & $3.2 \pm 0.7$ \\
 $Z_c^+(3900) \bar{D^0} {D^*}^+ $ & $2.5 \pm 0.3$ & $3.2 \pm 0.7$ \\ \botrule
\end{tabular} \label{ta3}}
\end{table}

From the results in Table 3 it is possible to evaluate the ratio
\begin{equation}
{\Gamma(Z_c(3900) \to D\bar{D}^*)\over
\Gamma(Z_c(3900) \to\pi J/\psi)}=0.22 \pm 0.12. 
\label{ratio}
\end{equation}

The QCDSR analysis performed in Ref.~[\refcite{gang3}] also supports the 
identification of $X(3872)$ and $Z_c^+(3900)$ as the $J^{PC} = 1^{++}$ and 
$J^{PC} = 1^{+-}$ diquark-antidiquark type tetraquark states, respectively.


\section{$Z^+_c (4025)$,  $Z^+_c (4020)$ and $Z^+_c (3885)$ : are they real ? }

Very recently  the BESIII Collaboration  reported the observation of 
other three charges states: $Z_c^+(4025)$ \cite{Ablikim:2013emm},
$Z_c^+(4020)$ \cite{Ablikim:2013wzq}   and $Z_c^+(3885)$ \cite{Ablikim:2013xfr}.

In the BESIII set-up a reaction $ e^+ e^− \rightarrow (D^* \bar{D^*})^{\pm}
 \pi^{\mp} $ was performed at $\sqrt{s} = 4.26$ GeV and a  
peak was  seen in the $ (D^*  \bar{D^*})^{\pm}  $ invariant mass distribution just
 about $10$ MeV above the threshold. The peak was identified as a new particle, 
the $Z_c^+(4025)$ \cite{Ablikim:2013emm}. 
The authors assume in the paper that the  $(D^*\bar{D^*})^{\pm}$
 pair is created in a S-wave and then the $Z_c^+(4025)$  
must have $J^P = 1^+$  to match, together with the pion, the quantum numbers 
$J^P = 1^-$ of the virtual photon from the $e^+ e^-$ pair. However, they also 
state that the experiment does not exclude other spin-parity assignments. 
Since the   $ (D^* \bar{D^*})^{\pm}  $  has charge, the isospin must be I = 1. 

In parallel with the experimental works many theoretical papers were devoted to 
understand these new states. In Ref.~[\refcite{hidalgo}], Heavy Quark Spin Symmetry
(HQSS) was used to make predictions for states containing one $D$ or $D^*$ and one 
$\bar{D}$ or $\bar{D^*}$. Assuming the $X(3872)$  to be $D \bar{D^*}$ 
molecule, the authors found a series of new hadronic molecules, including the 
$Z_c^+(3900)$ and the $Z_c^+(4025)$. They  would correspond to bound states (with 
uncertainties of about $50$ MeV in the binding) of $D\bar{D^*}$ and $D^*\bar{D^*}$ 
 respectively, with quantum numbers $I(J^{P}) = 1(1^{+})$.  Remarkably, even with 
uncertainties, these states always appear  in the bound region. In 
Refs.~[\refcite{chen,cui}], using QCD sum rules and assuming a
structure of $D^* \bar{D^*}$, the authors obtained a possible $I(J^P ) = 1(1^+)$ 
state compatible with the $Z_c^+(4025)$ albeit with around $250$ MeV uncertainty 
in the energy. Recently \cite{khetonn}, a study of the $D^* \bar{D^*}$ system has 
also been done within QCD sum rules, projecting the correlation function
on spin-parity $0^+$, $1^+$ and $2^+$. In the three cases a state with mass 
$3950 \pm 100$ MeV was found. The central value of the mass of these states is 
more in line with the results of Refs.~[\refcite{raquel,oset}], although with the error 
bar, they could as well be related to a resonance. In Ref.~[\refcite{sunzhu}] 
the new $Z_c$ states were investigated from a different perspective and, using 
pion exchange, a $D^* \bar{D^*}$ state with $I(J^P ) = 1(1^+)$ compatible with the
 $Z_c(4025)$ was obtained. One should note that the input used in this latter 
work is quite different from the one in Ref.~[\refcite{hidalgo}] since in HQSS the
pion exchange is subdominant. Finally, in Ref.~[\refcite{qtang}], using a 
tetraquark structure and QCDSR, a state with $I(J^P ) = 1(2^+)$ compatible with 
$Z_c(4025)$ was obtained, once again with a large error in the energy of $190$ MeV.
 In a different analysis,  in Ref.~[\refcite{matsuki}]  a pion and
the $D^* \bar{D^*}$  state are produced from the $X(4260)$ and the $D^* \bar{D^*}$
 state is left to interact, while the pion remains a spectator (initial 
single-pion emission mechanism). Although it is not mentioned
whether the $D^* \bar{D^*}$  interaction produces a resonance with certain quantum
 numbers,  the authors show that the mechanism can produce some enhancement in the
 $D^* \bar{D^*}$  invariant mass distribution just above threshold.

Bumps close to the threshold of a pair of particles should be treated with caution.
 Sometimes they are identified as new particles, but they can also be a reflection
 of a resonance below threshold. In  a similar reaction, $ e^+ e^- \rightarrow J/
\psi   (D \bar{D}) $, the Belle Collaboration reported \cite{pak} a bump close to 
the threshold in the  $  (D \bar{D}) $  invariant mass distribution, which was 
tentatively interpreted as a new resonance.
However, in Ref.~[\refcite{gamm08}] it was shown that the bump was better 
interpreted in terms of a      $  (D \bar{D}) $  molecular state, below the   
$  (D \bar{D}) $  threshold (the so called $X(3700)$). Similarly, in 
Ref.~[\refcite{tokhenno13}]   the $ \phi \,  \omega $ threshold peak measured
\cite{abli06}  in the $ J/\psi \rightarrow \gamma \phi \omega$  reaction was 
better interpreted as a signal of the $ f_0(1710)$ resonance, below the $\phi \,
\omega $ threshold, which couples strongly to  $\phi \,  \omega$.  Further 
examples of this phenomenon  may be found in Ref.~[\refcite{tokhenno14}]. In that 
work the theory of  $ D^* \bar{D^*}  $   interactions is reviewed and it 
is pointed out that a  $ (D^* \bar{D^*})  $  state with a mass above the 
threshold is very difficult to support. In particular, in Ref.~[\refcite{oset}] 
it was found that there is only one bound state of $ (D^* \bar{D^*})  $ 
in $I^G = 1^-$, with quantum numbers $J^{PC} = 2^{++}$  with a mass around 
$3990$ MeV and a width of about $100$ MeV. Both mass and width are compatible
with the reanalysis of data carried out in Ref.~[\refcite{tokhenno14}]. Therefore,
we can conclude that such $J^{P} = 2^{+}$ $D^{*}\bar{D}^*$ bound state provides a 
natural explanation for the state observed in  \cite{Ablikim:2013emm}.

An argument against the existence of a new resonance above the threshold is the 
fact that if the state were a $J^P = 1^+$  produced in S-wave, as assumed in the 
experimental work, it would easily decay into $J/\psi \pi$  exchanging a $D$ meson
 in the t-channel. This is also the decay channel of
the $Z_c(3900)$, which would then have the same quantum numbers as
the state claimed in Ref.~[\refcite{Ablikim:2013emm}]. However, while a peak is 
clearly seen in the $J/\psi \pi$ invariant mass distribution in the case of the 
$Z_c(3900)$, no trace of a peak is seen around $4025$ MeV in spite of using the 
same reaction and the same $e^+ e^- $ energy.

Less than a  month after the observation of the $Z_c^+(4025)$, the BESIII  
Collaboration reported the observation of the $Z_c^+(4020)$, a structure observed 
in the $h_c \pi^{\pm}$ mass spectrum \cite{Ablikim:2013wzq} . 
The difference between the parameters of 
this structure and the $Z_c^+(4025)$, observed in the  $ D^* \bar{D^*}  $  final 
state, is within $1.5 \, \sigma$ and it is not clear whether they are the 
same state or not.  The authors  do not find a significant signal for 
$Z_c^+(3900) \rightarrow h_c \pi^{\pm}$. 

Since the $Z_c^+(4025)$ and the $Z_c^+(4020)$ have almost the same mass and their 
quantum numbers were not yet accurately determined, we might think that 
they are, in fact, the same particle. Looking only at the most natural quantum 
numbers of the final states, the S-wave $D^* \bar{D^*}$ states have the quantum 
numbers  $J^{P} = 0^{+}$,  $1^{+}$ and $2^{+}$,  while the S-wave $h_c 
\pi^{\pm}$ states have the quantum numbers $J^{P} = 1^{-}$. Therefore the 
$Z_c^+(4025)$ and $Z_c^+(4020)$ would be  different particles. However, it is 
also possible to have a P-wave  $h_c \pi^{\pm}$  system with  quantum numbers 
$J^{P} = 0^{+}$ $1^{+}$ and $2^{+}$. In this case   the $Z_c^+(4025)$ 
and the $Z_c^+(4020)$ could be the same particle.

In the analysis presented in Ref.~[\refcite{gang1}], the author concluded that 
QCDSR do not support the picture of  $Z_c^+(4025)$ and  $Z_c^+(4020)$ 
as  diquark-antidiquark vector tetraquark states with $J^{P} = 1^{- }$. A 
short time later, in Ref.~[\refcite{gang2}]  the author concluded that, 
for these two states (treated as a single state), the QCDSR analysis supports the 
assignements  $J^{P} = 1^{+}$ and $J^{P} = 2^{+}$ in a diquark-antidiquark 
configuration.

Shortly after the observation of the $Z_c^+(4020)$ the same collaboration reported 
the measurement of the $Z_c^+(3885)$, a charged structure 
observed in the $(D \bar{D^*})\pm$ invariant mass distribution \cite{Ablikim:2013xfr}.  
The mass and width
 of this structure are $2 \sigma$  and $ 1 \sigma$, respectively,
below those of the $Z_c^+(3900)$.  The angular distribution of the  $\pi  Z_c(3885)
$ system favors the  $J^{P} = 1^{+}$ assignment and disfavors   
$J^{P} =  1^-$  or  $J^{P} =  0^-$.  Regarding the fact that this state could be
the $Z_c(3900)$, saw in a different decay channel, the only comment from the 
experimental side  is that if the $Z_c^+(3900)$ and $Z_c^+(3885)$ are the same 
state, then the ratio
\begin{equation}
{\Gamma(Z_c(3885) \to D\bar{D}^*)\over
\Gamma(Z_c(3900) \to\pi J/\psi)}=6.2 \pm 1.1 \pm 2.7
\label{bes}
\end{equation}
 is determined \cite{Ablikim:2013xfr}. 

Comparing the results in Eqs.~(\ref{ratio}) and (\ref{bes}) we can conclude
that, if the ratio in Eq.~(\ref{bes}) is confirmed, the states $Z_c^+(3900)$ and 
$Z_c^+(3885)$ are not the same state.

Here again, as in the case of the $Z_c^+(4025)$ discussed above, it is possible 
that the  $Z_c^+(3885)$) is not a real state but a manifestation of 
a resonance with a mass below the $(D \bar{D^*})$ threshold.  This point remains 
to be clarified. 


\section{Towards a new spectroscopy}

The proliferation of new charmonium states motivates attempts to group them into families. 
One possible way to organize some of the charmonium and bottomonium  new states  was
suggested in Ref.~[\refcite{Navarra:2011xa}] and it is summarized in Fig. 2. In 
this figure we present the charm and bottom spectra in the mass region of interest.
On the left (right) we show the charm (bottom) states with
their mass differences in MeV. The comparison between the two left lines with
the two lines on the right emphasizes the similarity between the spectra.
In the bottom of the second column  we have the  newly found $Z_c^+(3900)$.

The existence of a charged partner of the $X(3872)$ was first proposed in Ref.~[\refcite{maiani}]. A few 
years later\cite{maiani2} the same group proposed that the $Z^+(4430)$, observed 
by BELLE\cite{bellez},  would be the first radial excitation of the charged
 partner of the $X(3872)$. This suggestion was supported by  the fact that the mass
difference corresponding to a radial excitation in  the charmonium sector is
given by $M_{\Psi (2S)} - M_{\Psi (1S)}\sim 590$ MeV. This number is  close to
the mass difference $ M_{Z^+(4430)} - M_{X^+(3872)}\sim 560$ MeV. A similar 
connection between  $ Z^+(4430)$ and $Z_c^+(3900)$ was found in the 
hadro-charmonium approach\cite{voloshin2}, where the former is essentially a 
$\Psi^{\prime}$ embedded in light mesonic matter and the latter a $J/\psi$ also 
embedded in light mesonic matter. In Ref.~[\refcite{Navarra:2011xa}] this reasoning was extended to the
 bottom sector and it was 
conjectured that the $Z_b^+(10610)$, observed by the BELLE collaboration in 
Ref.~[\refcite{bellezb}],  might be a radial excitation of
an yet unmeasured $X_b^+$, predicted in Ref.~[\refcite{x3872}]. The observation of 
$Z_c^+(3990)$ gives support to this conjecture and should motivate  new 
experimental searches of this bottom charged state and its neutral partner, the 
only missing states in the diagram.
\begin{figure}[h]
\centerline{\epsfig{figure=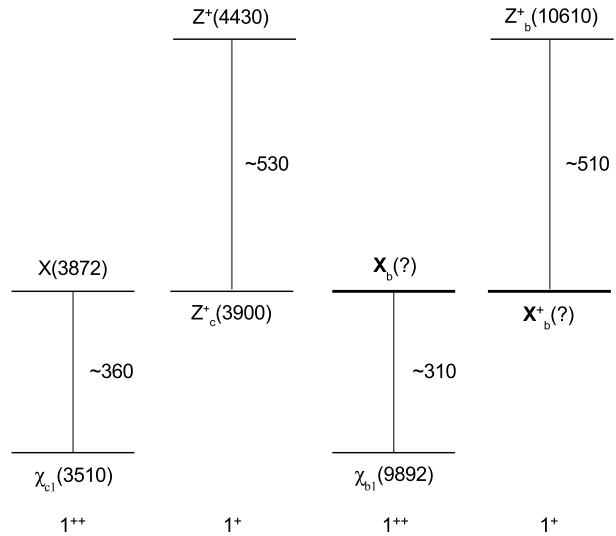,height=80mm}}
\caption{Charm and bottom energy levels}
\label{fig2}
\end{figure}


\section{Conclusion}

The most important message from the experimental program carried out at Belle and 
BESIII is that definitely there is something really 
new happening in the charmonium spectroscopy. This  started in 2003 with the 
measurement of the $X(3872)$,  which  has a very robust experimental 
signature and has been neasured by many different groups. The $X(3872)$ is 
electrically neutral and hence its multiquark nature was not 
clear from the beginning. Five years later, in 2008, the observation of 
$Z^+(4430)$, $Z_1^+(4050)$ and $Z_2^+(4250)$ would have been the proof of 
the existence of multiquark configurations in the charmonium sector. However the 
non-confirmation of these measurements rendered this claim weak. Another five
years later, in 2013, the confirmation of the observation of the $Z^+(4430)$ 
together with the measurements of the $Z^+_ c(3900)$ (which was 
measured by BESIII and confirmed by other groups) and also of the $Z_c^+(4025)$, 
$Z_c^+(4025)$ and $Z_c^+(3885)$, reinforced our belief that we are 
observing multiquark states. What has to be done next? From the experimental side 
it is necessary to determine unambiguosly the quantum numbers 
of all these states and eliminate the suspicion that they are mere threshold 
effects and not real particles. As suggested in Ref.~[\refcite{tokhenno14}], this 
can be done performing an energy scan in the $e^+ e^-$ reactions. Moreover, a 
more refined analysis will allow us to determine whether all these states are 
really different.  From the theoretical side  its necessary to focus on the 
calculation of the decay widths in all the different approaches, since, as we 
have discussed, the masses are easily obtained by different methods and they 
are not sufficient to discriminate between different theoretial models.  If our 
present picture of these states survives all these tests and 
improvements, we will have found multiquark states. This is in itself very 
interesting! Whether meson molecules, tetraquarks or hadrocharmonium, 
these are novel objects which will induce a small revolution in our understanding 
of hadrons.

\section*{Acknowledgments}

The authors are grateful to E. Oset, A. Martinez Torres, K. Kemchandani, J. 
Nieves, M.P. Valderrama and A. Ozpineci for fruitful discussions. 
This work has been supported by CNPq, FAPESP and USP (NAP-75).

\end{document}